\documentstyle[12pt,axocolor,cite]{article}
\def\RedViolet{} 
\def\Black{}
\def\Blue{}

\def\RawSienna{}
\def\beq{\begin{equation}}
\def\eeq{\end{equation}}
\def\bea{\begin{eqnarray}}
\def\eea{\end{eqnarray}}
\def\bq{\begin{quote}}
\def\eq{\end{quote}}

\def\bib{\bibitem}
\def\be{\begin{equation}}
\def\ee{\end{equation}}
\def\barr{\begin{array}}
\def\earr{\end{array}}
\def\dis{\displaystyle}

\def\eg{ {\em e.g.}}

\def\etal{ {\em et al.}}
\def\ie{ {\em i.e.}}

\def\im{{\rm Im}}

\def\mev{\: {\rm MeV} }
\def\gev{\: {\rm GeV} }

\def\ra{\rightarrow}

\def\gappeq{\mathrel{\rlap
{\raise.5ex\hbox{$>$}}
{\lower.5ex\hbox{$\sim$}}}}

\def\lappeq{\mathrel{\rlap{\raise.5ex\hbox{$<$}}
{\lower.5ex\hbox{$\sim$}}}}
\def\Hwk{{\cal H}_{\rm wk}}
\def\cG{{\cal G}}
\def\cV{{\cal V}}
\def\cA#1{ {\cal A}_{#1} }
\def\cD{{\cal D}}
\def\cR#1{ {\cal R}_{#1} }
\def\tm#1{ {\tilde m}_{#1} }
\begin{document}
\setcounter{page}{0}
\thispagestyle{empty}
\setcounter{footnote}{0}
\renewcommand{\thefootnote}{\fnsymbol{footnote}}

\pagestyle{empty}
\begin{flushright}
		MRI-PHY/P980443\\[1.5ex]
                CERN-TH/98-76\\[1.5ex]
                {\large \tt hep-ph/9804300} \\
\end{flushright}

\vskip 25pt

\begin{center}
\advance\baselineskip by 10pt

{\Large \sc {\Blue
Estimates of Long-distance Contributions\\ to the}
{\Black 
$B_s \rightarrow \gamma \gamma$ }
{\Blue Decay} }\\

\Black
\advance\baselineskip by -10pt
\vspace*{1cm} 
{\bf Debajyoti Choudhury}$^{a)}$\footnote{debchou@mri.ernet.in}
 and
{\bf John Ellis}$^{b)}$\footnote{John.Ellis@cern.ch} \\

\vspace{13pt}

{\em $^{a)}$  Mehta Research Institute
        of Mathematics and Mathematical Physics, \\
      Chhatnag Road,       Jhusi, 
      Allahabad -- 211019, India
}

{\em $^{b)}$ 
        Theory Division, CERN, CH 1211 Geneva 23, 
        Switzerland
}\\[2ex]

\vspace{30pt}
{\bf ABSTRACT} \\
\end{center}
\vspace*{5mm}

\begin{quotation}
\noindent
{\RawSienna 
We present first calculations of new long-distance contributions
to $B_s \rightarrow \gamma \gamma$ decay due to intermediate
$D_s$ and $D_s^*$ meson states. The relevant $\gamma$ vertices
are estimated using charge couplings and transition moment
couplings. Within our uncertainties, we find that these
long-distance contributions could be comparable to the known
short-distance contributions. Since they have different
Cabibbo-Kobayashi-Maskawa matrix-element factors, there may be
an interesting possibility of observing CP violation in this decay.}
\end{quotation}
 
\Black
\vspace*{50pt} 
\noindent

\begin{flushleft} CERN-TH/98-76
\\
April 1998
\end{flushleft}
\vfill\eject

\setcounter{page}{1}
\pagestyle{plain}
\setcounter{footnote}{0}
\renewcommand{\thefootnote}{\arabic{footnote}}

A new era of experiments on rare $B$ decays is about to dawn. The
large data sets already obtained at LEP and CESR will be dwarfed
by those provided by the $e^+ e^-$ $B$ factories, and by the hadron
experiments HERA-B, CDF, D0 and LHC-B. Among the rare $B$ decays
with particularly clean experimental signatures is $B_s \rightarrow \gamma
\gamma$, whose branching ratio presently has the experimental upper limit
{\RedViolet
${\cal B}(B_s \rightarrow \gamma\gamma) 
< 1.48 \times 10^{-4}$}~\Black\cite{L3}.
Higher-sensitivity
measurements of $B_s$ decays are not among the highest priorities of
the $e^+ e^-$ $B$ factories, whose physics programmes are focussed
initially on searches for CP violation in $B_d$ decays, but the hadron
experiments cannot avoid
being sensitive to ${\cal B}(B_s \rightarrow \gamma \gamma)$ to
levels several orders of magnitude below the present experimental upper
limit~\cite{L3}.

The lowest-order short-distance contributions to the 
$B_s \rightarrow \gamma\gamma$ decay arise from two sets of graphs:
($i$) box diagrams, and ($ii$) triangle diagrams with an external 
photon leg. These have been calculated~\cite{LLY,Sim_Wyl,Kalin}, 
and yield, for 
$m_t \approx 175 \gev$,\footnote{Over a wide range of the 
               top-quark mass, the partial width grows
	       linearly with $m_t$~\protect\cite{Kalin}.}
{\RedViolet
${\cal B}(B_s \rightarrow \gamma\gamma) 
\sim 3.8 \times 10^{-7}$
},
\Black 
the next generation 
of experiments. Interestingly, the branching fraction can be 
enhanced substantially in extensions of the Standard Model such 
as a generic 2-Higgs scenario~\cite{2Higgs}. Emboldened by the fact that 
the short-distance QCD corrections are not too 
big~\cite{CLY,2Higgs,SRC_Yao}, one may then even attempt to use such data
as may be forthcoming as probes for new physics.

However, it is not immediately obvious that this decay is dominated
by such short-distance contributions. 
Estimates
have been made of some long-distance contributions to the decay
amplitudes, for example via intermediate charmonium states $J/\psi$ and
$\psi'$~\cite{long-dist}.
These were found to be very small, indicating that the short-distance
contribution might be correct in
order of magnitude. But, before concluding this to be the case,
one must examine other possible long-distance contributions.

\begin{figure}[htb]
\input{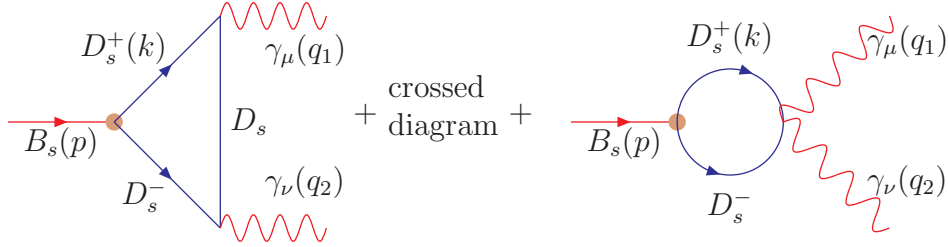}
   \caption{\em One-loop contribution to the
	       $B_s \ra \gamma \gamma$
	      amplitude due to $D_s$ mesons alone. 
	      }
	\label{fig:pure-d}
\end{figure}

\begin{figure}[htb]
\input{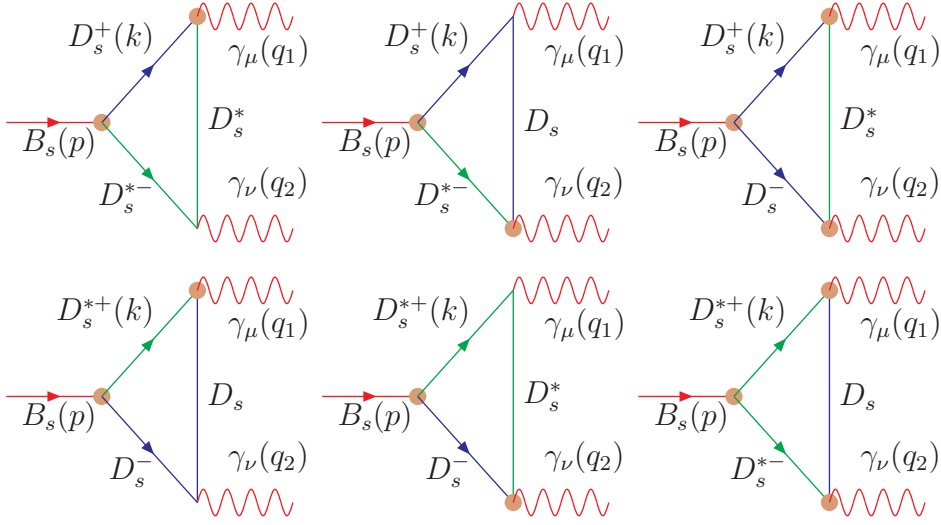}
   \caption{\em One-loop contribution to the
	       $B_s \ra \gamma \gamma$
	      amplitude involving the $D_s^\ast D_s \gamma$ vertex.
	      The crossed diagrams are not shown. 
	      }
	\label{fig:mixed}
\end{figure}

In this paper we study such contributions due to
intermediate $D_s$ and $D^*_s$  states via the diagrams
shown in Figures~\ref{fig:pure-d} and \ref{fig:mixed}. 
These include loops of $D_s$ mesons alone,
loops of $D_s^*$ mesons alone, and diagrams involving radiative $D_s
\rightarrow D_s^*$ transitions. 
The vertices are estimated using  data on $B \rightarrow DD, DD^*$
and $D^* D^*$ decays, the electromagnetic charge couplings of the $D_s$
and $D_s^*$, and phenomenological estimates of the $D_s^* \rightarrow D_s
+ \gamma$ decay rate for the inelastic transition-matrix element.
The imaginary part of the diagrams
can be calculated easily while the determination 
of the real part requires the use of dispersion relations.

We find that these long-distance contributions to $B_s \rightarrow
\gamma\gamma$ decay may be comparable to the short-distance
contributions calculated previously, within the considerable
uncertainties inherent to the phenomenological inputs we use.
We find no evidence that ${\cal B}(B_s \rightarrow \gamma\gamma)$
decay should occur at a rate very close to the present
experimental upper limit, but even this possibility cannot be ruled out.
Since the short- and long-distance contributions 
have different Cabibbo-Kobayashi-Maskawa matrix-element
factors, the fact that they may be of
comparable magnitudes offers the interesting possibility of
observing CP violation in this decay mode. However, a detailed
exploration of this possibility lies beyond the scope of this paper.

\vspace*{3ex}
At the quark level, the operators responsible for the 
${\cal B}(B_s \rightarrow \gamma\gamma)$ decay 
are $\bar b \gamma_5 s F_{\mu \nu} F^{\mu \nu}$  and 
$\bar b \gamma_5 s F_{\mu \nu} \widetilde F^{\mu \nu}$, where $F$
is the electromagnetic fileld strength tensor, and $\widetilde F$ 
is its dual. The matrix element for the decay 
$B_s \ra \gamma(q_{1\mu}) \gamma(q_{2\nu})$ 
can then be parametrized as 
\be
\barr{rcl}
\RawSienna
    {\cal M}    & = & \dis \alpha \cG 
                          \left( \cR1 S_{\mu \nu} + i \cR2 P_{\mu \nu} 
                          \right) \\
\Black
    S_{\mu \nu} & \equiv & 
                  \dis q_{1\nu} q_{2 \mu} - q_1 \cdot q_2 g_{\mu \nu}\\
    P_{\mu \nu} & \equiv & \dis \epsilon_{\mu \nu \alpha \beta}
                        q_1^{\alpha} q_{2}^{\beta} \\
    \cG   & = & \dis \frac{G_F}{\sqrt{2} } V_{cb} V_{cs}^\ast
\earr
	\label{hadronic-me}
\ee
where $\cR1$ and $\cR2$ are the yet-to-be-determined hadronic matrix
elements, and 
$(\alpha \cG)$ has been factored out in anticipation of the results
to be derived later in the text.
The partial width is then given by
\be
	\Gamma(B_s \ra \gamma \gamma) = \frac{(\alpha \cG)^2}{64 \pi} 
					  m_B^3
					\left( |\cR1|^2 + |\cR2|^2 \right)
         \ .
     \label{partial_width}
\ee
Since $\Gamma (B_s) \approx 5 \times 10^{-10} \mev$, we have, 
for $|V_{cb}| = 0.04$, 
\be
{\RawSienna
   Br (B_s \ra \gamma \gamma)  \approx
          10^{-7} 
               \; \left(  \left| \frac{\cR1}{100 \mev} \right|^2
                       + \left| \frac{\cR2}{100 \mev} \right|^2
                  \right)
}
\Black
   \ .
	\label{brfrac}
\ee
It can be argued that the long-distance contributions should 
be dominated by two-meson intermediate states. Now, 
the quark-level transitions with the least Cabibbo suppression are 
$b \ra c \bar c s $ and $b \ra c \bar u d $. The second process
obviously leads to very dissimilar mesons which may rescatter into 
two photons only through higher-order terms.
Consequently, we concentrate on the $b \ra c \bar c s $
case, with the corresponding weak Hamiltonian being given by 
\be
\barr{rcl}
    \Hwk & = & \cG J_\mu j^\mu 
         \\[1.2ex]
    J_\mu & = & \bar c  \gamma_\mu (1 - \gamma_5) b 
         \\[1.2ex]
    j^\mu & = & \bar s  \gamma_\mu (1 - \gamma_5) c  \  ,
\earr
	\label{weakHamilt}
\ee
where $V$ denotes the CKM matrix. 
The relevant intermediate states are then $D_s^+ D_s^-$, 
$D_s^{\ast +} D_s^{\ast -}$, 
$D_s^{\ast +} D_s^-$, etc.\footnote{Since the contributions due to 
			charmonia have already been 
			investigated~\protect\cite{long-dist}, 
			we shall not consider these here.}. 
To calculate the 
$B_s \ra \gamma \gamma$ decay amplitude exactly, 
we need to determine the corresponding matrix elements of 
(\ref{weakHamilt}), a non-trivial task. However, as a first 
approximation, we may assume naive 
factorization\footnote{Based on experience elsewhere, we may hope that
                   the associated errors are $< {\cal O}(10\%)$.}. 
The $H_{\rm wk}$ matrix elements are then expressible in terms of 
simpler quantities given by
\be
\barr{rcl}
  \langle D_s^-(q) | j^\mu | 0 \rangle & = & -i f q^\mu 
         \\[1.2ex]
  \langle D_s^{\ast -}(k,\epsilon) | j^\mu | 0 \rangle 
         & = & m_\ast f_\ast \epsilon^{\ast \mu}
         \\[1.2ex]
  \langle D_s^-(k) | J_\mu | B_s(p) \rangle & = & 0
         \\[1.2ex]
  \langle D_s^{\ast -}(k,\epsilon) | J_\mu | B_s(p) \rangle & = & 0
         \\[1.2ex]
  \langle D_s^+(k) | J_\mu | B_s(p) \rangle & = & 
               f^+(q^2) (p + k)_\mu  + f^-(q^2) q_\mu
         \\[1.2ex]
  \langle D_s^{\ast +}(k,\epsilon) | J_\mu | B_s(p) \rangle 
         & = & \dis \frac{\cV}{m_\ast} 
                    \epsilon_{\mu \nu \alpha \beta} \epsilon^{\ast \nu} 
                 p^\alpha k^\beta 
         \\[1ex]
         & + & \dis 
              \frac{i}{m_\ast}
                \left[ m_B^2\cA1 \epsilon^{\ast}_\nu 
                        - \epsilon^{\ast} \cdot p 
                           \left\{ \cA2 (k + p)_\mu 
                                   + \frac{m^2_\ast}{q^2} \cA3 q_\mu 
                           \right\}
                 \right]  
        \  ,
\earr
	\label{matrixelem}
\ee
where $ q = p - k $ in the last two terms, and $m_\ast \equiv m_{D^\ast}$.
The quantities
$\cV$ and $\cA{i}$ are related to the usual form factors through
\be
\barr{rcl}
    \cV & = & \dis \frac{2 m_\ast}{m_B + m_\ast} 
                V_{B_s D_s^\ast}(q^2)
            \\[1.3ex]
    \cA1 & = & \dis \frac{m_\ast(m_B + m_\ast)}{m_B^2}
                A^1_{B_s D_s^\ast}(q^2)
            \\[1.3ex]
    \cA2 & = & \dis \frac{m_\ast}{m_B + m_\ast} 
               A^2_{B_s D_s^\ast}(q^2)
            \\[1.5ex]
    \cA3 & = & \dis 2 \left[
                     A^3_{B_s D_s^\ast}(q^2) - A^0_{B_s D_s^\ast}(q^2) 
                      \right]
	\ .
\earr
	\label{formfac}
\ee
Finally, we have
\be
\barr{rcl}
  \langle D_s^+ (k) \; D_s^-(q) | \Hwk | B_s(p) \rangle 
               & = & -i \cG f \; \left[ (p^2 - k^2) f^+(q^2) 
                                          + q^2 f^-(q^2) \right]
         \\[1.2ex]
  \langle D_s^{\ast +} (k) \; D_s^{\ast -} (q) | \Hwk | B_s(p) \rangle 
       & = & \cG f_\ast \;
                 \epsilon^{\ast \mu}_{+} \epsilon^{\ast \nu}_{-} \;
            \left[ \cV(q^2) \epsilon^{\nu \mu \alpha \beta} p_\alpha k_\beta 
                \right. 
                  \\[1ex]
       & & \dis \hspace*{6em}
                 \left.
                 + i \left\{ \cA1(q^2) p^2 g^{\mu \nu} 
                                - 2 \cA2(q^2) p^\mu p^\nu \right\}
            \right] 
         \\[1.2ex]
  \langle D_s^{\ast +}(k,\epsilon) \; D_s^-(q) | \Hwk | B_s(p) \rangle 
         & = & \dis \frac{\cG f}{m_\ast}
               \; \epsilon^\ast \cdot p \;
                \left[ p^2 \cA1(q^2) - (p^2 - k^2) \cA2 - m^2_\ast \cA3 (q^2)
                \right]
         \\[1.4ex]
  \langle D_s^+(k) D_s^{\ast -} (q,\epsilon) | \Hwk | B_s(p) \rangle 
         & = & 2 \cG m_\ast f_\ast f^+(q^2) 
                 \; \epsilon^{\ast} \cdot p 
      \ .
\earr
	\label{BDD-vertex}
\ee
Within the factorization approximation,the relations in 
(\ref{BDD-vertex}) can be interpreted as a 
parametrization of the $B D_s^{(\ast)} D_s^{(\ast)}$ vertices. 

The form factors themselves are nonperturbative quantities. 
Apart from lattice calculations~\cite{Lattice}, the best arenas
for determining these are phenomenological approaches such as the 
BSW model~\cite{BSW} or heavy-quark effective theory~\cite{HQET}. 
Some of these parameters are also measured experimentally~\cite{f_Ds}, 
although a somewhat large spread persists in the data. Typically,
though, we have $f, f_\ast \sim 200 \mev$,  
$f_\pm, \cV \sim 0.6$ 
and $\cA{1,2} \sim 0.25$. On the other hand, $\cA3$ is
very small. 

As a next step, we need to determine the meson-photon couplings. Since 
we are dealing here with charged mesons, the simplest course is to 
assume minimal substitution. This then fixes the charge coupling uniquely
for each angular-momentum state of the ($c \bar s$) system. Of 
course, the fact that these mesons are not fundamental particles
means that we should ideally discuss a series of 
form factors, including charge radii as well as higher moments. As
a first approximation, however, we neglect such aspects of the 
meson-photon coupling. 

The one-loop contribution due to the $D_s$ alone can then 
be expressed in terms of the diagrams in Fig.~\ref{fig:pure-d}. 
Computing these diagrams, the first one gives
\[
\barr{rcl}
G_{\mu \nu}^{(1)} & = & \dis 
       \int \frac{d^4 k}{(2 \pi)^4} (-i \cG ) f \;
	          \left[(p^2 - k^2) f_+ + (p - k)^2 f_- \right]
	          \frac{i}{k^2 - m^2} (i e )
		   \left[2 k + q_1 \right]_\mu
                \\[1.5ex]
	& & \dis \hspace*{1.5em} 
		\frac{i}{(k-q)^2 - m^2} (i e )
		   \left[2 k - q_1 - p \right]_\mu
		\frac{i}{(k-p)^2 - m^2} 
        \ ,
\earr
\]
whilst the third gives
\[
G_{\mu \nu}^{(3)}  = 
       \int \frac{d^4 k}{(2 \pi)^4} (-i \cG ) f \;
	          \left[(p^2 - k^2) f_+ + (p - k)^2 f_- \right]
	          \frac{i}{k^2 - m^2} 
		\frac{i}{(k-p)^2 - m^2} 
		(2 i e^2 g_{\mu \nu})
     \ .
\]
In each of the above, the form-factors $f_\pm$ are to be evaluated at 
$(p - k)^2$. 

We note that an exact calculation of the above integrals needs 
knowledge of the momentum dependence of the form factors. 
Furthermore, the integrals are formally divergent, and 
to calculate the real parts we would 
need to consider a cutoff scale that sets the limit of validity of 
a theory with `fundamental' meson fields. Rather than attempt this, 
we take recourse to the optical theorem and calculate only the 
absorptive parts of the diagrams above. On using the Cutkosky rules, 
the sum of the three diagrams reduces to
\be
\barr{rcl}
\im( G_{\mu \nu}^{(D_s)} ) 
	        & = &  2 \alpha \cG \tm1 f \; 
			\left[ (1 - \tm1) f^+(m^2) + \tm1 f^-(m^2) \right]
	I_{111} S_{\mu \nu}
	\\[1.5ex]
    I_{ijk} & = & \dis \ln \: 
                   \frac{1 + 2 \tm{k} - \tm{i} - \tm{j} + \lambda_{ij} }
                        {1 + 2 \tm{k} - \tm{i} - \tm{j} - \lambda_{ij} }
	\\[2ex]
    \lambda_{ij} & = & \dis \left[ ( 1 - \tm{i} - \tm{j})^2 
                                   - 4 \tm{i} \tm{j}
                            \right]^{1/2} 
	\ .
\earr
	\label{pure-d}
\ee
Here $    \tm{i} = m_i^2 / m_B^2 $ with $m_1 = m \equiv m_{D_s}$ and
$    m_2  =  m_\ast \equiv m_{D_s^\ast} $. 
The form of (\ref{pure-d}) is a testimonial to the fact that the 
three diagrams of Fig.~\ref{fig:pure-d} together form a gauge-invariant
set. Thus, if $D_s$ were the only meson that could contribute, 
$\cR{1,2}$ would be determined by (\ref{pure-d}), and
\be
\barr{rcl}
\RedViolet
\im \cR1 (D_s)
  & = & \dis 0.39\; f \; \left[ f^+(m^2) +  0.16 f^-(m^2) \right]
  \\[1.5ex]
\im \cR2 (D_s) 
  & = & 0 
\Black
\earr
	\label{R_D}
\ee
The CP-violating form factor is thus identically zero. 

What about the real parts of the amplitudes $\cR1$?
These may be computed using dispersion relations~\cite{K_to_gg_vector},
and, for the case in hand, are seen to be smaller than the 
imaginary parts. 
Substituting for the form factors in the above and 
using (\ref{brfrac}), we then have 
\be
\Blue
 Br (B_s \buildrel 2 D_s \over {\longrightarrow}
 \gamma \gamma) \sim 2.5 \times 10^{-8} 
\Black
	\ .
\ee
This, by itself, is small compared to the short-distance 
contribution~\cite{CLY}, though the interference term can 
be significant. 

What about the other long-distance effects? 
Of the ones calculated in the literature, 
$B_s \ra \phi \gamma \ra \gamma \gamma $ 
has an amplitude somewhat smaller than the one calculated here, 
whilst $B_s \ra J/\psi \phi \ra \gamma \gamma $ is seen to 
contribute only at the 1\% level~\cite{long-dist}.
We are thus in a situation where the 
$2 D_s$ intermediate state may, in fact, give the largest 
long-distance contribution to the decay amplitude. 
It is interesting, at this stage, to compare 
the $B_s \ra \gamma \gamma$ case with 
that for $K_S \ra \gamma \gamma$ decay. Whereas there it is 
the $2 \pi$ intermediate state that dominates the decay 
amplitude~\cite{K_to_gg}, the analogous contribution to the 
$B_s$ decay falls well short of the short-distance amplitude.

We now turn our attention to the next higher state that can 
contribute to this process, namely the $D_s^\ast$. 
As long as we neglect any $D_s^\ast D_s \gamma$ coupling, the additional
diagrams involving the  $D_s^\ast$ field are exact analogues of those in 
Fig.~\ref{fig:pure-d}, and lead to 
\be
\barr{rcl}
\im \cR1 (D_s^\ast)      &= & \dis 
             \frac{f_\ast }{8 \tm2^2}
           \left[ 
               \cA1(m_\ast^2)
                  \left\{ \left( 1 - 4 \tm2 \right) \lambda_{22}
                        + \tm2 \left(1 - 12 \tm2 + 48 \tm2^2 \right) 
                              I_{222}
                  \right\} 
            \right.            \\[2.5ex]
       & & \dis \hspace*{3.5em}
           \left. 
               - \cA2(m_\ast^2)
                  \left\{ \left( \tm2 - 5 \right) \lambda_{22}
                        + \tm2 \left( 1 - 10 \tm2 + 32 \tm2^2 \right) 
                          I_{222}
                  \right\}
             \right]
        \\[2.5ex]
    & = & \RedViolet \dis f_\ast 
                 \left[ 1.57 \cA1(m_\ast^2)
                                            + 15.3 \cA2(m_\ast^2) \right]
                     \\[3.5ex]
\Black
\im \cR2 (D_s^\ast)
    &  = & \dis
        \frac{f_\ast \cV(m_\ast^2)}{8 \tm2}
          \left\{ \left( 12 \tm2 - 1 \right) \lambda_{22}
                + \left( 4 \tm2 - 32 \tm2^2 \right) I_{222}
          \right\} 
                     \\[2.5ex]
     & = & \RedViolet \dis 0.13 \; f_\ast \; \cV(m_\ast^2) 
\Black
\earr
	\label{R_Dstar}
\ee
Two new features confront us. One is that the $CP$-violating amplitude is
non-zero, though
small. More relevantly for our present purpose, we find 
${\rm Im} \cR1(D_s^\ast)
\sim 800 \mev$, on the basis of which
(\ref{brfrac}) then gives us
\be
\Blue
 Br (B_s \buildrel 2 D_s^\ast \over {\longrightarrow}
 \gamma \gamma) \sim 6.5 \times 10^{-6} 
	\label{brfrac_dst}
\Black
	\ .
\ee
This, of course, is much larger than the short-distance contribution, 
and suggests that this particular decay should be observable at the 
very first run of the hadronic machines! 

A few objections could possibly be raised against this
conclusion. For one thing, the validity 
of the factorization approximation as applied to decays 
into two vector mesons is not apparent. Thus, instead of using 
(\ref{matrixelem}), 
it would be preferable to use the data to parametrize the 
$B_s D_s^\ast \bar D_s^\ast$ vertex. Unfortunately, though, this 
exclusive mode is undetected so far. Once it is measured, it 
will be a straightforward task to reexpress our result as 
a prediction of the ratio of the two decay modes.

Secondly, we have, until now, neglected quite a few possible 
contributions to the process. For example, the contributions due to 
the higher excited states could in principle
be comparable in magnitude, and might interfere destructively, thus 
reducing the total long-distance contribution. However, in the absence
of extensive data on these states, it is almost impossible 
to estimate such contributions to any degree of reliability. 
We can only hope that this lack of information does not invalidate 
the results presented here.

Finally, there is another class of contribution that we have neglected
so far, involving
the $D_s^\ast D_s \gamma$ vertex. Whilst this transition moment
can, in principle, be calculated within a given model for the mesons,
it is easier to work in terms of an effective Lagrangian, the relevant
part of which can be expressed as 
\be
\RawSienna
	{\cal L}_{\rm eff} = i \: \frac{e \cD}{m_\ast} 
\epsilon_{\rho \sigma \alpha \beta}
                              q^\alpha p^\beta 
\Black
   \ .
     \label{dst-d-gamma}
\ee
We shall assume that $\cD$ is real, \ie\, there is no absorptive part 
associated with this vertex.
The new contributions are given by the diagrams of Fig.~\ref{fig:mixed},
along with the crossed ones, and the resultant shifts 
$\delta \cR{1,2}$ given by
\[
\barr{rcl}
4 \tm2 \: \im \:\delta \cR1 & = & \dis 
         |\cD|^2  f  \; 
           \left[ (1 - \tm1) f^+(m^2) + \tm1 f^-(m^2) \right]
             \left\{\lambda_{11} - \tm2 I_{112}
             \right\}
           \\[1.5ex]
  & - & \dis 
          | \cD |^2 f_\ast  \;
             \Bigg[ 
                \cA1 (m_\ast^2) 
	         \left\{ \lambda_{12} - \tm1 I_{221}
                 \right\}
               \\[1.5ex]
       &  & \dis \hspace*{4.8em} 
             \left.
               + \cA2 (m_\ast^2) 
	         \left\{ 
                     \left( \tm2 - \tm1 - 0.5 
                        \right) \lambda_{12} 
                      + \left( (\tm2 - \tm1)^2 + \tm1 
                           \right) I_{221}
                 \right\}
              \right]
\\[2ex]
8 \tm2 \: \im \delta \cR2 & = & \dis 
      \cD f_\ast  \; f^+(m_\ast^2) 
        \Big\{2 (\tm1^3 - \tm1^2 - 4 \tm1^2 \tm2 + 2 \tm1 \tm2 
		   + 5 \tm1 \tm2^2 - 2 \tm2^3 + 3 \tm2^2 - \tm2 ) I_{122}
           \\[1.3ex]
          & & \dis\hspace*{6em} 
		+ 8 \tm1 \tm2 I_{121} 
		+  3 (\tm1 - \tm1^2 + \tm2^2 - 3 \tm2 ) \lambda_{12}
         \Big\}
 \\[1.9ex]
   & + & \dis
         \cD  f
             \left[ \cA1 (m^2) + (\tm2 - 1) \cA2 (m^2) 
                            - \tm2 \cA3 (m^2) \right]
	\\
    & & \dis \hspace*{5em} 
		\Bigg\{ 4 \tm1 I_{211} + (1 - 3 \tm1) I_{112}
		     + \left( 2 \tm2 - \tm1 - 3 - \frac{\tm1 - \tm1^2}{\tm2}
			    \right) \lambda_{12}
         \Bigg\}
\\[1.9ex]
       & + & \dis 
         | \cD |^2 f_\ast
                        \cV (m_\ast^2)
         \left\{ \left( \tm2^2 - \tm1^2 \right) 
			    I_{221} 
		+ \left( \tm2 + \tm1 - 0.5 \right) \lambda_{12} 
	 \right\} 
     \  .
\earr
\]
Numerically, then, 
\be
\barr{rcl}
\RedViolet
  \im \:\delta \cR1   & = & \dis 
       | \cD |^2 \;
        \left\{
           0.077 \: f \: \left[ f^+(m^2) +  0.16 f^-(m^2) \right]
           + f_\ast \:
               \left[ - 0.67 \cA1(m_\ast^2) + 0.15 \cA2(m_\ast^2)
                             \right]
	\right\}
\\[3.5ex]
 \im \: \delta \cR2 & = & \dis
      0.15 \; \cD  f \;  
             \left[ \cA1 (m^2) - 0.845 \cA2 (m^2) 
                            - 0.155 \cA3 (m^2) \right]
       \\[1.5ex]
    & - & 
      0.098 \; \cV(m_\ast^2) \; f_\ast \; | \cD |^2  
     - 0.12 \: \dis \cD f_\ast \; f^+(m_\ast^2) 
\Black
\earr
	\label{deltar}
\ee
One may, in principle, estimate $\cD$ from the partial decay width of 
$D_s^\ast$:
\be \dis
\Gamma (D_s^\ast \ra D_s \gamma) 
       = \frac{\alpha}{24} |\cD|^2 \; m_\ast 
		\left( 1 - \frac{m^2}{m_\ast^2} \right)^3
       = ( 1.46 \times 10^{-3} \mev) \; |\cD|^2
    \label{dst_decay}
\ee
Unfortunately, this decay mode is not yet well measured. All we know 
is that $\Gamma (D_s^\ast) < 1.9 \mev$ and that $(D_s + \gamma)$ 
and $(D_s + \pi)$ are the only decay modes seen. This implies that
\[
\RawSienna
	|\cD|^2 < 1300 \; Br(D_s^\ast \ra D_s \gamma) 
\Black
      \ .
\]
A branching ratio of even 1\% could then lead to an order 
of magnitude change in $\Gamma(B_s \ra \gamma \gamma)$!
However, a stricter bound for $\cD$ can be obtained 
if one relates it to the corresponding form factor
for the $S = 0$ charm meson. Noting that 
$\Gamma (D^\ast) < 0.131 \mev$
and $Br (D^{\ast +} \ra D^+ \gamma)  \approx 1.7\%$~\cite{Dst_Dgam}, 
a relation analogous to (\ref{dst_decay}) leads to 
\[
\RawSienna
	| \cD_{D^\ast D \gamma} |^2 < 0.95  
\Black
	\ .
\]
A value of $\cD$ of this order obviously cannot negate the conclusions of 
(\ref{brfrac_dst}). Examining 
(\ref{deltar}) closely, we see 
that the contribution due to a pair of on-shell $D_s$ exchanging 
a $D_s^\ast$ is actually much smaller than that of (\ref{R_D}).
This result is similar again to 
the case of $K_S \to \gamma \gamma$ decay~\cite{K_to_gg_vector}, where 
the analogous contributions due to the flavour-octet vector mesons 
are small. The other diagrams, where at least one $D_s^\ast$ is
on shell, can, however, compete with the $2 D_s$ contribution. 

To conclude, we have estimated the long-distance contributions to the 
$B_s \ra \gamma \gamma$ decay arising from charmed-meson 
intermediate states. We find that the $2 D_s$ contribution, by itself, 
is larger than the other long-distance contributions calculated 
hitherto in the literature~\cite{long-dist}. It is, however, 
still smaller 
than the short-distance amplitude. More interestingly, the 
$2 D_s^\ast $ contribution is much larger, and could enhance the 
branching fraction by more than an order of magnitude. This 
would lead to a very striking signal at the hadronic $B$ factories.
Moreover, if it is indeed comparable to the short-distance amplitude,
the different Cabibbo-Kobayashi-Maskawa structures might offer interesting
prospects 
for observing CP violation.

DC acknowledges useful discussions with Ahmed Ali, Tariq Aziz, 
Leo Stodolsky and York-Peng Yao.

\end{document}